\documentclass[english,aps,reprint,superscriptaddress]{revtex4}
\usepackage[T1]{fontenc}
\usepackage[latin9]{inputenc}
\pdfoutput=1
\usepackage{amsmath}
\usepackage{amssymb}
\usepackage{graphicx}
\usepackage{subscript}

\makeatletter
\@ifundefined{textcolor}{}
{%
 \definecolor{BLACK}{gray}{0}
 \definecolor{WHITE}{gray}{1}
 \definecolor{RED}{rgb}{1,0,0}
 \definecolor{GREEN}{rgb}{0,1,0}
 \definecolor{BLUE}{rgb}{0,0,1}
 \definecolor{CYAN}{cmyk}{1,0,0,0}
 \definecolor{MAGENTA}{cmyk}{0,1,0,0}
 \definecolor{YELLOW}{cmyk}{0,0,1,0}
}

\usepackage{pslatex}

\usepackage{babel}

\makeatother

\usepackage{babel}
\begin{document}

\title{Radiation Pressure Force from Optical Cycling on a Polyatomic Molecule}

\author{Ivan Kozyryev}

\email{ivan@cua.harvard.edu}

\affiliation{Harvard-MIT Center for Ultracold Atoms, Cambridge, MA 02138, USA}

\affiliation{Department of Physics, Harvard University, Cambridge, MA 02138, USA}

\author{Louis Baum}

\affiliation{Harvard-MIT Center for Ultracold Atoms, Cambridge, MA 02138, USA}

\affiliation{Department of Physics, Harvard University, Cambridge, MA 02138, USA}

\author{Kyle Matsuda}

\affiliation{Harvard-MIT Center for Ultracold Atoms, Cambridge, MA 02138, USA}

\affiliation{Department of Physics, Harvard University, Cambridge, MA 02138, USA}

\author{Boerge Hemmerling }

\affiliation{Harvard-MIT Center for Ultracold Atoms, Cambridge, MA 02138, USA}

\affiliation{Department of Physics, Harvard University, Cambridge, MA 02138, USA}

\affiliation{Present address: Department of Physics, University of California,
Berkeley, CA 94720, USA}

\author{John M. Doyle}

\affiliation{Harvard-MIT Center for Ultracold Atoms, Cambridge, MA 02138, USA}

\affiliation{Department of Physics, Harvard University, Cambridge, MA 02138, USA}
\begin{abstract}
We demonstrate multiple photon cycling and radiative force deflection
on the triatomic free radical strontium monohydroxide (SrOH). Optical
cycling is achieved on SrOH in a cryogenic buffer-gas beam by employing
the rotationally closed $P\left(N''=1\right)$ branch of the vibronic
transition $\tilde{X}^{2}\Sigma^{+}\left(000\right)\leftrightarrow\tilde{A}^{2}\Pi_{1/2}\left(000\right)$.
A single repumping laser excites the Sr-O stretching vibrational mode,
and photon cycling of the molecule deflects the SrOH beam by an angle
of $0.2^{\circ}$ via scattering of $\sim100$ photons per molecule.
This approach can be used for direct laser cooling of SrOH and more
complex, isoelectronic species. 
\end{abstract}
\maketitle

\section{\label{sec:Introduction}Introduction}

The use of laser radiation to control and cool external and internal
degrees of freedom for neutral atoms \cite{chu1998nobel}, molecules
\cite{hamamda2015ro,Shuman2010}, microspheres \cite{kiesel2013cavity,asenbaum2013cavity},
and micromechanical membranes \cite{peterson2016laser,chan2011laser}
has revolutionized atomic, molecular, and optical physics. The powerful
techniques of laser cooling and trapping using light scattering forces
for atoms led to breakthroughs in both fundamental and applied sciences,
including detailed studies of diverse degenerate quantum gases \cite{Greif2016,bakr2010probing},
creation of novel frequency standards \cite{ludlow2015optical}, precision
measurements of fundamental constants \cite{fixler2007atom,clade2006determination},
and advances in quantum information processing \cite{maller2015rydberg,jessen2004quantum}.
Beyond atoms, cold and ultracold molecules beckon with promising applications
in controlled chemistry \cite{krems2008cold}, many-body physics \cite{carr2009cold,wang2006quantum},
and quantum science \cite{andre2006coherent,rabl2006hybrid}.

Following proposals by Di Rosa \cite{di2004laser} and Stuhl et al.
\cite{stuhl2008magneto}, as well as initial experimental results
by Shuman et al. \cite{Shuman2009}, laser cooling has been successfully
applied to a few diatomic molecules (SrF \cite{Shuman2010}, YO \cite{Hummon2013}
and CaF \cite{zhelyazkova2014laser}) and led to magneto-optical trapping
of SrF at sub-millikelvin temperatures \cite{norrgard2015sub}. Recently,
Isaev and Berger \cite{isaev2015polyatomic} identified promising
polyatomic molecules for direct Doppler cooling based on Franck-Condon
(FC) factor calculations. While a nonresonant dipole force was used
to deflect a beam of CS\textsubscript{2} \cite{stapelfeldt1997deflection}
and optical pumping led to rotational cooling of CH\textsubscript{3}F
\cite{glockner2015rotational}, there has been no demonstration of
the radiative force via optical cycling for a species with more than
two atoms.

Polyatomic molecules are more difficult to manipulate than atoms and
diatomic molecules because they possess additional constituents and
their concomitant additional rotational and vibrational degrees of
freedom. Partially because of their increased complexity, cold dense
samples of molecules with three or more atoms offer unique capabilities
for exploring interdisciplinary frontiers in physics, chemistry and
even biology. Precise control over polyatomic molecules could lead
to applications in astrophysics \cite{herbst2009complex}, quantum
simulation \cite{wall2015realizing} and computation \cite{tesch2002quantum},
fundamental physics \cite{kozlov2013linear,kozlov2013microwave},
and chemistry \cite{sabbah2007understanding}. Study of parity violation
in biomolecular chirality \cite{quack2008high} - which plays a fundamental
role in molecular biology \cite{quack2002important} - necessarily
requires polyatomic molecules consisting of at least four atoms.

In recent years, a number of different experimental tools were developed
for controlling neutral gas-phase polyatomic molecules, including
Stark deceleration followed by electric trapping \cite{bethlem2000electrostatic},
pulsed magnetic slowing \cite{momose2013manipulation}, buffer-gas
cooling \cite{patterson2015slow}, rotating centrifugal slowing \cite{chervenkov2014continuous},
and optical Stark deceleration \cite{fulton2004optical}. However,
with the exception of optoelectrical Sisyphus cooling \cite{Zeppenfeld2012,prehn2016optoelectrical},
the lowest temperatures reached were around 1 K \cite{lemeshko2013manipulation}.
While association of ultracold atoms into diatomic molecules has allowed
creation of nearly degenerate bialkali molecular gases in the singlet
ground state \cite{ni2008high,park2015ultracold}, using similar methods
to create more complex polyatomics appears challenging at the moment. 

In this paper, we demonstrate optical cycling and the radiation pressure
force on triatomic strontium monohydroxide (SrOH). The structure of
SrOH is more complicated compared to previously laser cooled diatomic
molecules: it contains three vibrational modes \cite{presunka1995laser},
including degenerate bending vibrations with no direct analog in diatomic
molecules. Additionally, the Renner-Teller effect, which is absent
in diatomics, further complicates molecular structure \cite{brown1983vibronic}.
Our experimental results demonstrate that despite the significantly
increased complexity associated with even a simple polyatomic molecule
like SrOH, optical cycling on a quasi-closed transition can be applied
with technically straightforward modifications. In particular, for
this species (and those isoelectronic to it, to be discussed later
in this paper), the additional degrees of freedom do not affect photon
cycling up to the level of hundreds of photons, more than enough to
implement deflection through radiative force. To reach the level of
thousands of scattered photons (necessary to cool to the millikelvin
regime), very similar techniques can be used, for example in SrOH
by repumping the excited bending mode.

\section{Experimental details}

\begin{figure}[h]
\begin{centering}
\includegraphics[height=7cm]{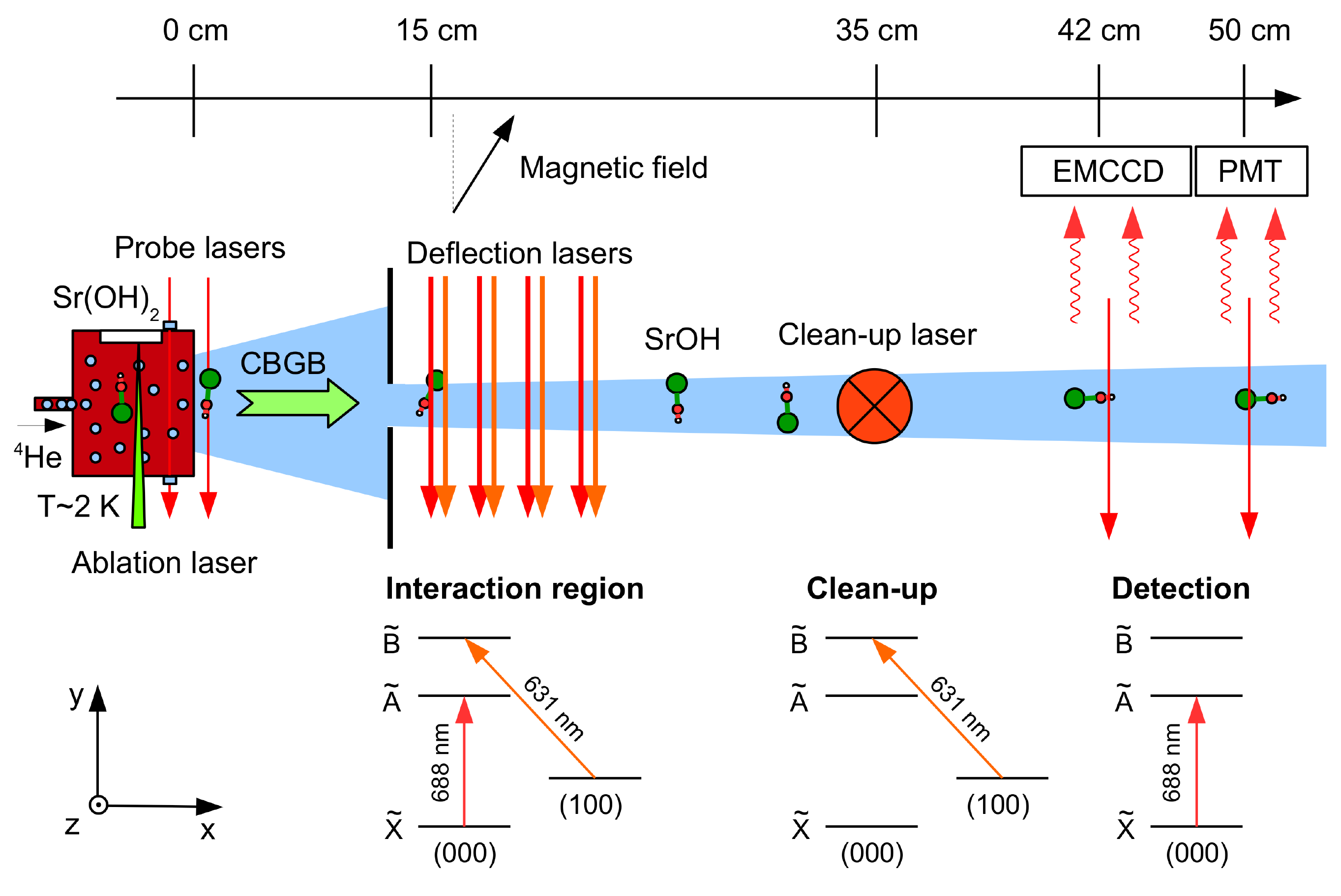} 
\par\end{centering}

\protect\protect\caption{\label{fig:Schematic-of-the-apparatus}Schematic of the experimental
setup (not to scale). A cryogenic beam of SrOH is produced using laser
ablation of a pressed Sr(OH)\protect\protect\textsubscript{2} target
followed by buffer-gas cooling with $\sim2\ {\rm K}$ helium gas.
Transverse lasers, resonant with $P\left(N''=1\right)$ line of the
$\tilde{X}^{2}\Sigma^{+}\left(000\right)\rightarrow\tilde{A}^{2}\Pi_{1/2}\left(000\right)$
and $\tilde{X}^{2}\Sigma^{+}\left(100\right)\rightarrow\tilde{B}^{2}\Sigma^{+}\left(000\right)$
electronic transitions, interact with the collimated molecular beam
in order to apply radiation pressure force. In order to remix the
dark magnetic sub-levels, a magnetic field at an angle of $60^{\circ}$
is applied in the interaction region. Molecules remaining in the excited
vibrational level of the electronic ground state are optically pumped
back into the ground vibrational level using $\tilde{X}\rightarrow\tilde{B}$
off-diagonal excitation. The spatial profile of the molecular beam
is imaged on the electron multiplying charge-coupled device (EMCCD)
camera and the time-of-flight data is collected on the photomultiplier
tube (PMT). The vibrational quantum numbers $\left(v_{1}v_{2}v_{3}\right)$
correspond to the Sr$\leftrightarrow$OH stretching ($v_{1}$), Sr-O-H
bending ($v_{2}$), and SrO$\leftrightarrow$H stretching ($v_{3}$)
vibrational modes. }
\end{figure}

A schematic diagram of the experimental apparatus is shown in figure
\ref{fig:Schematic-of-the-apparatus}. Gas-phase SrOH is produced
by laser ablation of solid Sr(OH)\textsubscript{2} placed inside
a cryogenic cell maintained at a temperature of $\sim2\ {\rm K}$.
The study of SrOH buffer-gas cooling dynamics, as well as measurements
of its momentum transfer and inelastic cross sections with helium,
were previously performed \cite{kozyryev2015collisional}. Detailed
descriptions of the cryogenic and vacuum chambers used in this experiment
have been provided elsewhere \cite{hemmerling2013buffer}. Briefly,
SrOH molecules entrained in helium buffer gas flowing into the cell
at a rate of 6 sccm (standard cubic centimeters per minute) are extracted
into a beam through a $5\ {\rm mm}$ aperture. This cryogenic buffer-gas
beam (CBGB) \cite{hutzler2012buffer} contains approximately $10^{9}$
molecules in the first excited rotational level ($N=1$) in a pulse
about 5 ms long. The forward velocity of the SrOH beam is $130\pm20\ {\rm m/s}$
and its transverse velocity spread is $\pm15\ {\rm m/s}$. A rectangular
$2\times2$ mm slit situated 15 cm away from the cell aperture collimates
the beam. Deflection of the molecules is achieved by applying laser
light perpendicular to the beam's flight path. To increase the interaction
time, several laser beams are applied in a series. The deflection
laser beams originate from a single-mode fiber containing two colors
- 688 nm for driving the $\tilde{X}^{2}\Sigma^{+}\left(000\right)\rightarrow\tilde{A}^{2}\Pi_{1/2}\left(000\right)$
transition and 631 nm for driving the $\tilde{X}^{2}\Sigma^{+}\left(100\right)\rightarrow\tilde{B}^{2}\Sigma^{+}\left(000\right)$
transition. The exact scheme is described in more detail in section
\ref{sec:Results-and-discussion}. Each color contains two frequency
components separated by $\sim110\ {\rm MHz}$ to address $P_{11}\left(J''=1.5\right)$
and $^{P}Q_{12}\left(J''=0.5\right)$ lines of the spin-rotation (SR)
splitting. Each dual-color beam has FWHM diameter of 1.8 mm and contains
$50\ {\rm mW}$ of total laser power. In order to create multiple
passes (to maximize molecule deflection), the same beam is circulated
around the vacuum chamber. The light is generated using injection-locked
laser diodes seeded by external-cavity diode lasers in the Littrow
configuration \cite{cunyun2004tunable}. In order to destabilize the
dark states created during the cycling process \cite{berkeland2002destabilization},
we apply a magnetic field of a few gauss at an angle of $60^{\circ}$
to the polarization plane $xz$. 

In the ``clean-up'' region, we repump all the molecular population
from the excited vibrational level $\tilde{X}\left(100\right)$ back
to the ground state in order to increase the signal in the detection
region. The spatial profile of the molecular beam is extracted by
imaging the laser-induced fluorescence (LIF) from a transverse retroreflected
laser beam on an EMCCD camera. The laser beam addresses both SR components
of the $P\left(N''=1\right)$ line for the $\tilde{X}^{2}\Sigma^{+}\left(000\right)\rightarrow\tilde{A}^{2}\Pi_{1/2}\left(000\right)$
transition with $\sim1{\rm {\rm mW}}$ each. In a similar laser configuration,
time-of-flight data is recorded by collecting the LIF on a PMT 8 cm
downstream.

\section{\label{sec:Results-and-discussion}Results and discussion}

\begin{figure}[h]
\begin{centering}
\includegraphics[width=7cm]{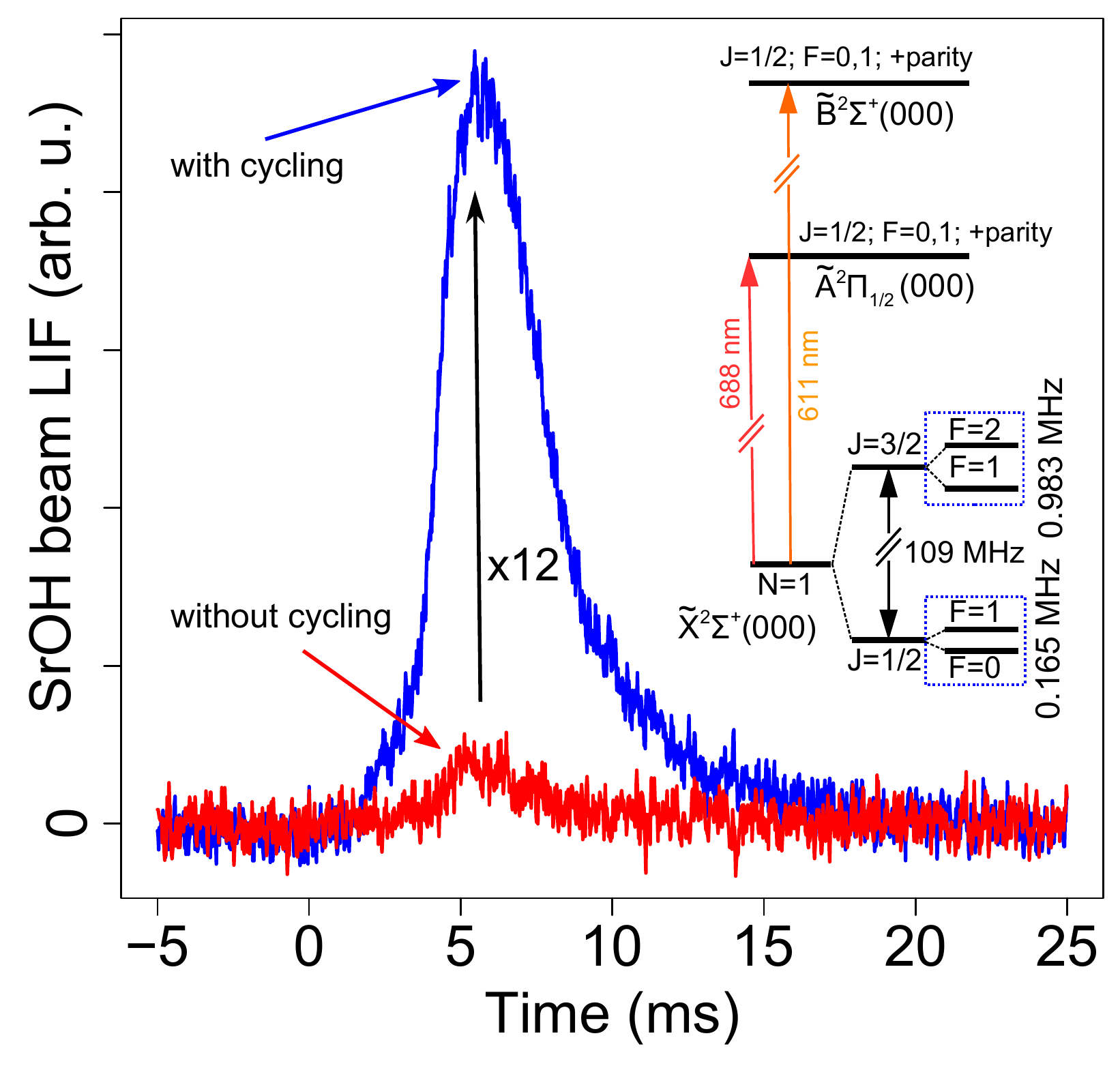} 
\par\end{centering}

\protect\protect\caption{\label{fig:Laser-induced-fluorescence-enhancement}Laser induced fluorescence
(LIF) increase due to photon cycling. The red trace shows the SrOH
beam signal in the first excited rotational level ($N=1$) when only
one of the spin-rotation lines is addressed with the 688 nm laser
resonant with the $P\left(N''=1\right)$ line of the $\tilde{X}^{2}\Sigma^{+}\rightarrow\tilde{A}^{2}\Pi_{1/2}$
electronic transition. More than an order of magnitude LIF increase
is observed (in blue) when two laser frequencies separated by $\sim110\ {\rm MHz}$
excite both of the spin-rotation lines. Increased fluorescence corresponds
to the scattering of approximately 24 photons per each molecule limited
by the decay to the dark vibrational level. The inset diagram shows
relevant rotational, fine and hyperfine structure levels of SrOH.
Rotationally closed excitations on the $\tilde{X}-\tilde{A}$ and
$\tilde{X}-\tilde{B}$ electronic transition are shown with red and
orange upward arrows, correspondingly. The unresolved hyperfine splittings
(grouped in dashed squares) have previously been measured \cite{fletcher1993molecular}
and are smaller than the natural linewidth of the $\tilde{X}-\tilde{A}$
electronic transition \cite{nakagawa1983high}. }
\end{figure}

\begin{figure}[h]
\begin{centering}
\includegraphics[width=7cm]{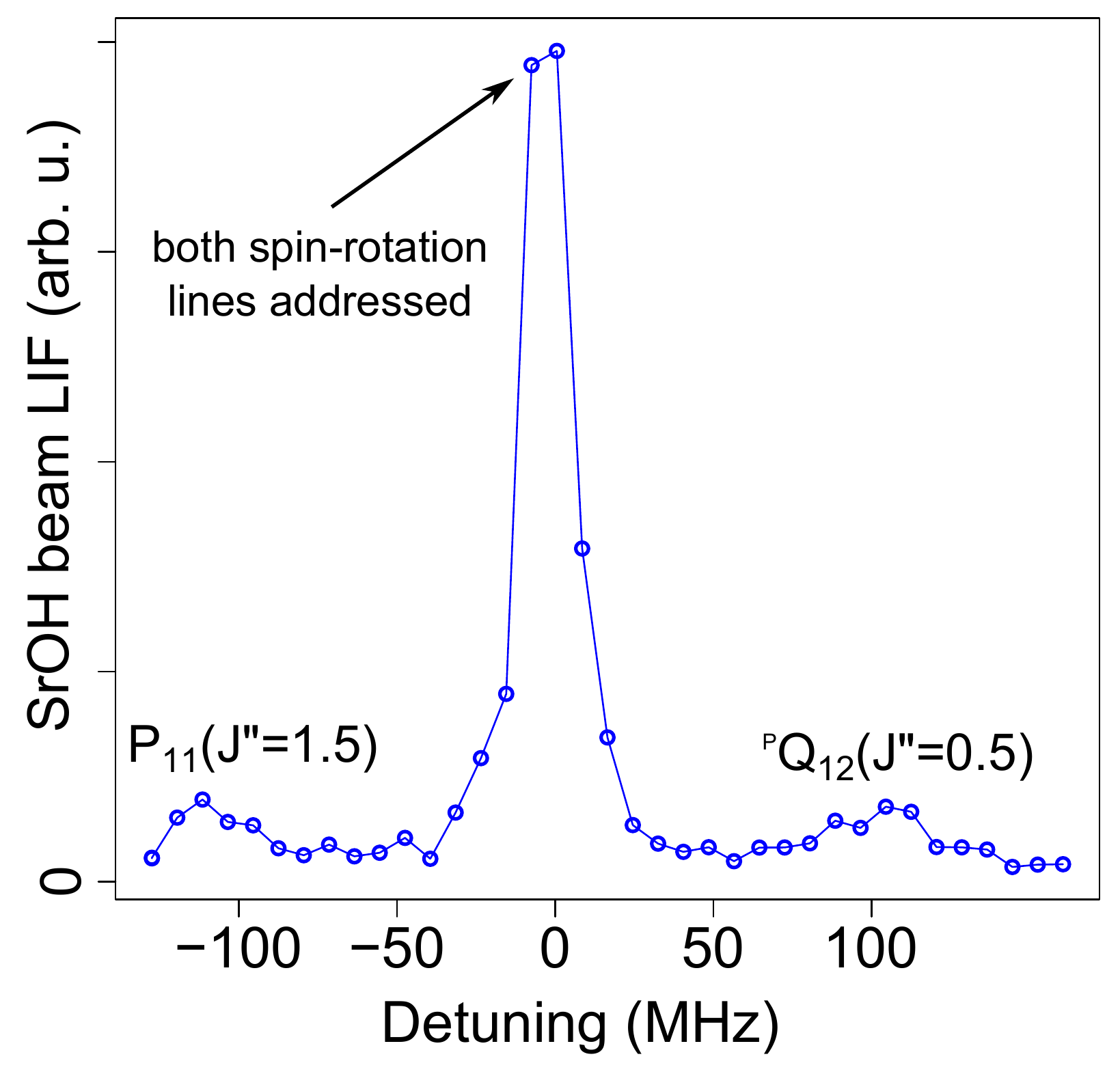} 
\par\end{centering}

\protect\protect\caption{\label{fig:Fluorescence-spectrum-with-cycling}Fluorescence spectrum
with photon cycling. A three-peak spectrum is observed upon scanning
of two laser frequencies separated by $\sim110\ {\rm MHz}$ through
the two spin-rotation components of the $P\left(N''=1\right)$ line
for the $\tilde{X}^{2}\Sigma^{+}\left(000\right)-\tilde{A}^{2}\Pi_{1/2}\left(000\right)$
transition. Due to the photon cycling process, collected fluorescence
with both spin-rotation lines addressed is an order of magnitude larger
than the sum of the individual signals.}
\end{figure}

The rotationally closed electronic transitions used in this work are
shown in the inset of figure \ref{fig:Laser-induced-fluorescence-enhancement}.
We chose the lowest frequency $\tilde{X}-\tilde{A}$ electronic excitation
at 688 nm for optical cycling in the experiment because it can be
addressed with all solid-state lasers. The diagonal FC factors of
the $\tilde{X}-\tilde{A}$ band \cite{brazier1985laser,kozyryev2015collisional}
allow for scattering multiple photons before decaying to excited vibrational
levels. Following a previously developed scheme for diatomics \cite{Shuman2009,zhelyazkova2014laser,Hummon2013,hemmerling2016CaF},
we address the molecules in the first excited rotational level on
the $P\left(N''=1\right)$ line. Because of the rotational selection
rules, molecules return to the same rotational state after the cycle
$N''=1\leftrightarrow N'=0$. Hyperfine splittings in SrOH are below
the natural linewidth of the electronic transition \cite{fletcher1993molecular}.
The observed fluorescence enhancement due to photon cycling is demonstrated
in figure \ref{fig:Laser-induced-fluorescence-enhancement}. The red
trace shows the molecular beam signal without photon cycling when
only the $P_{11}\left(J''=1.5\right)$ line is addressed and, therefore,
on average, the molecules will scatter approximately two photons before
decaying to the dark SR component. Adding the second laser frequency
to excite the $^{P}Q_{12}\left(J''=0.5\right)$ transition, we see
more than an order of magnitude increase in the LIF, which corresponds
to about 24 scattered photons per molecule. We model the number of
photon scattering events before a decay to the dark vibrational level
$\tilde{X}\left(100\right)$ as a geometric distribution with probability
of success $p$, identified as the ``off-diagonal'' FC factor $f_{\tilde{A}\left(000\right)\rightarrow\tilde{X}\left(100\right)}$.
Since the expected value of the geometric probability distribution
is $1/p$ \cite{rice2006mathematical}, on average, $1/f_{\tilde{A}\left(000\right)\rightarrow\tilde{X}\left(100\right)}\approx25$
photons are scattered per molecule before the loss to the dark vibrational
level, which agrees with our observations. Figure \ref{fig:Fluorescence-spectrum-with-cycling}
shows the scan of two laser frequencies together, where the spin-rotation
components manifest themselves as peaks detuned by $\sim110\ {\rm MHz}$
from the center peak. The cycling LIF is larger than the combined
signal for both of the SR components alone, indicating photon cycling
in the molecular system.

\begin{figure}[h]
\begin{centering}
\includegraphics[width=7cm]{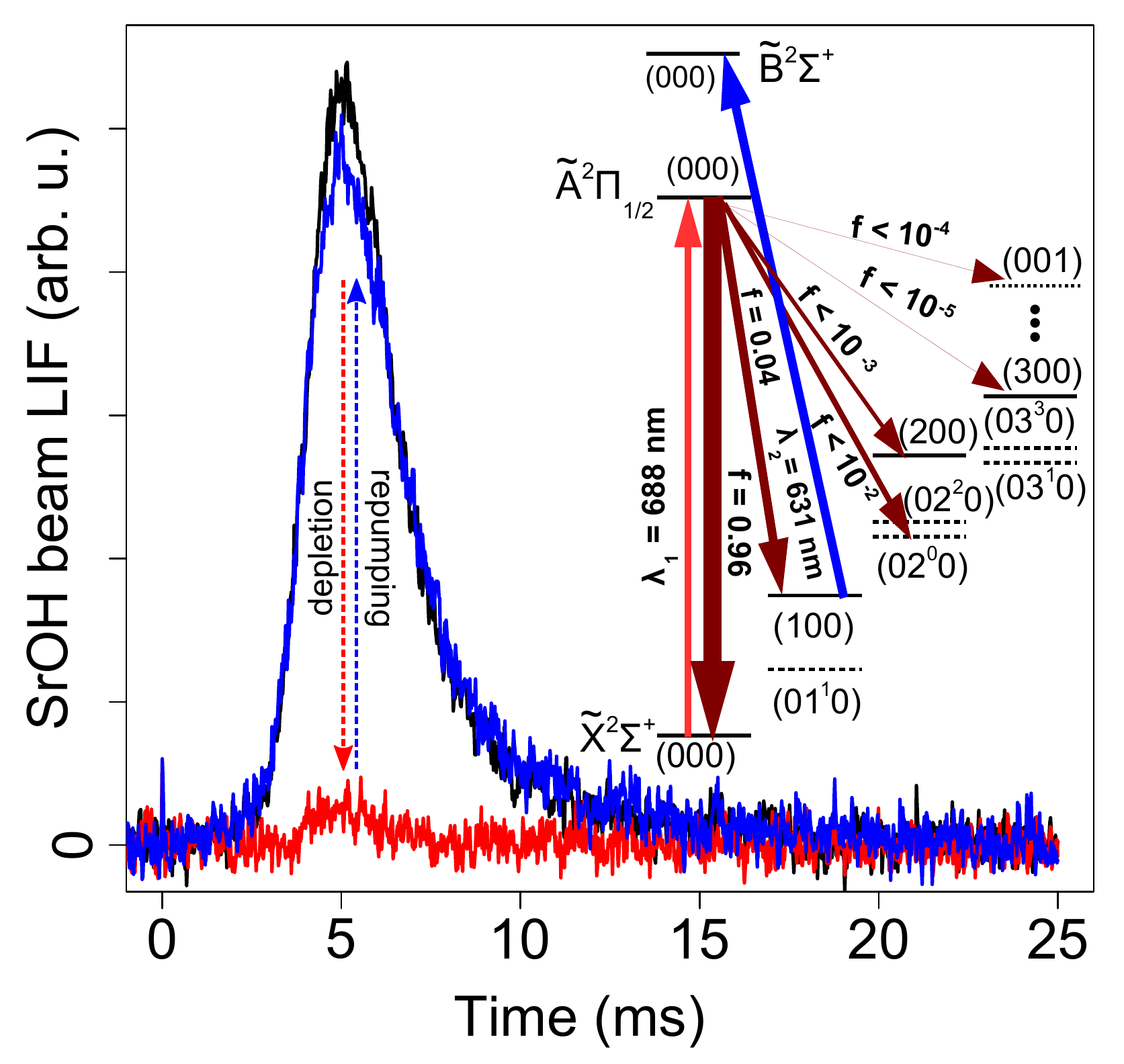} 
\par\end{centering}

\protect\protect\caption{\label{fig:Cycling-between-vibrational-levels}Cycling between vibrational
levels. The molecular beam signal (black curve) is depleted (red curve)
by the main cycling laser at 688 nm (light red, inset) due to off-diagonal
vibrational decay to the excited vibrational mode (100). The signal
is recovered (blue curve) by the application of the 631 nm repumping
laser (blue, inset) in the clean-up region. The inset diagram shows
the details of the vibrational structure of SrOH relevant to the optical
cycling scheme. The energies of the excited vibrational levels in
the electronic ground state have been previously measured \cite{presunka1995laser}.
The main cycling ($\lambda_{1}$) and repump ($\lambda_{2}$) lasers
are indicated with the upward arrows, while the spontaneous decay
channels in the Born-Oppenheimer approximation are shown with downward
dark-red arrows. Thicker decay lines correspond to stronger transitions
with the corresponding Franck-Condon factors ($f$) indicated. The
values and bounds for FC factors are taken from \cite{SteimleISMS}.
The rotational states are not depicted at this scale. The superscript
$l$ next to the bending mode vibrational quantum number ($v_{2}^{l}$)
indicates the projection of the vibrational angular momentum on the
internuclear axis.}
\end{figure}

The experimentally relevant vibrational structure of SrOH is depicted
in the inset of figure \ref{fig:Cycling-between-vibrational-levels}.
Upon electronic excitation, $96\%$ of the molecules return to the
vibrational ground state while $4\%$ decay to the excited Sr-O stretching
mode (100). We repump these molecules via the $\tilde{B}$ state using
the 631 nm laser, as shown in the diagram. Figure \ref{fig:Cycling-between-vibrational-levels}
depicts cycling between the (000) and (100) vibrational levels of
the ground electronic state $\tilde{X}$. By applying light in the
interaction region, we pump all $\tilde{X}\left(000\right)$ molecules
(in black) into the excited vibrational mode $\tilde{X}\left(100\right)$
after scattering of $\sim25$ photons, which is indicated by the depleted
beam profile (in red). Application of the repumping beam returns the
molecules to the ground vibrational level and we recover the fluorescence
signal (in blue). Cycling between vibrational levels indicates that
the dominant vibrational loss mechanism in the molecular system is
to the excited Sr-O stretching mode $\tilde{X}\left(100\right)$.
It is estimated \cite{SteimleISMS,kozyryev2015collisional} that the
molecules will scatter about 1,000 photons before they decay to the
second excited Sr-O stretching mode $\tilde{X}\left(200\right)$. 

As previously mentioned in section \ref{sec:Introduction}, in addition
to Sr-O stretching, SrOH contains two other vibrational modes that
can limit the photon cycling process. Because the vibrational angular
momentum selection rule $\triangle l=0$ allows only for specific
decays to the excited bending mode vibrations from the $\tilde{A}\left(000\right)$
state \cite{herzberg1966molecular}, the dominant loss channel in
the bending mode is to $\tilde{X}\left(02^{0}0\right)$ energy level
with $l=0$. SrOH molecules will scatter at least 10,000 photons before
decaying to the excited O-H stretching mode $\tilde{X}\left(001\right)$
\cite{OberlanderPhDthesis}. Therefore, laser cooling can be effectively
performed without the need for the O-H stretching mode repumping laser.

\begin{figure}[h]
\begin{centering}
\includegraphics[width=7cm]{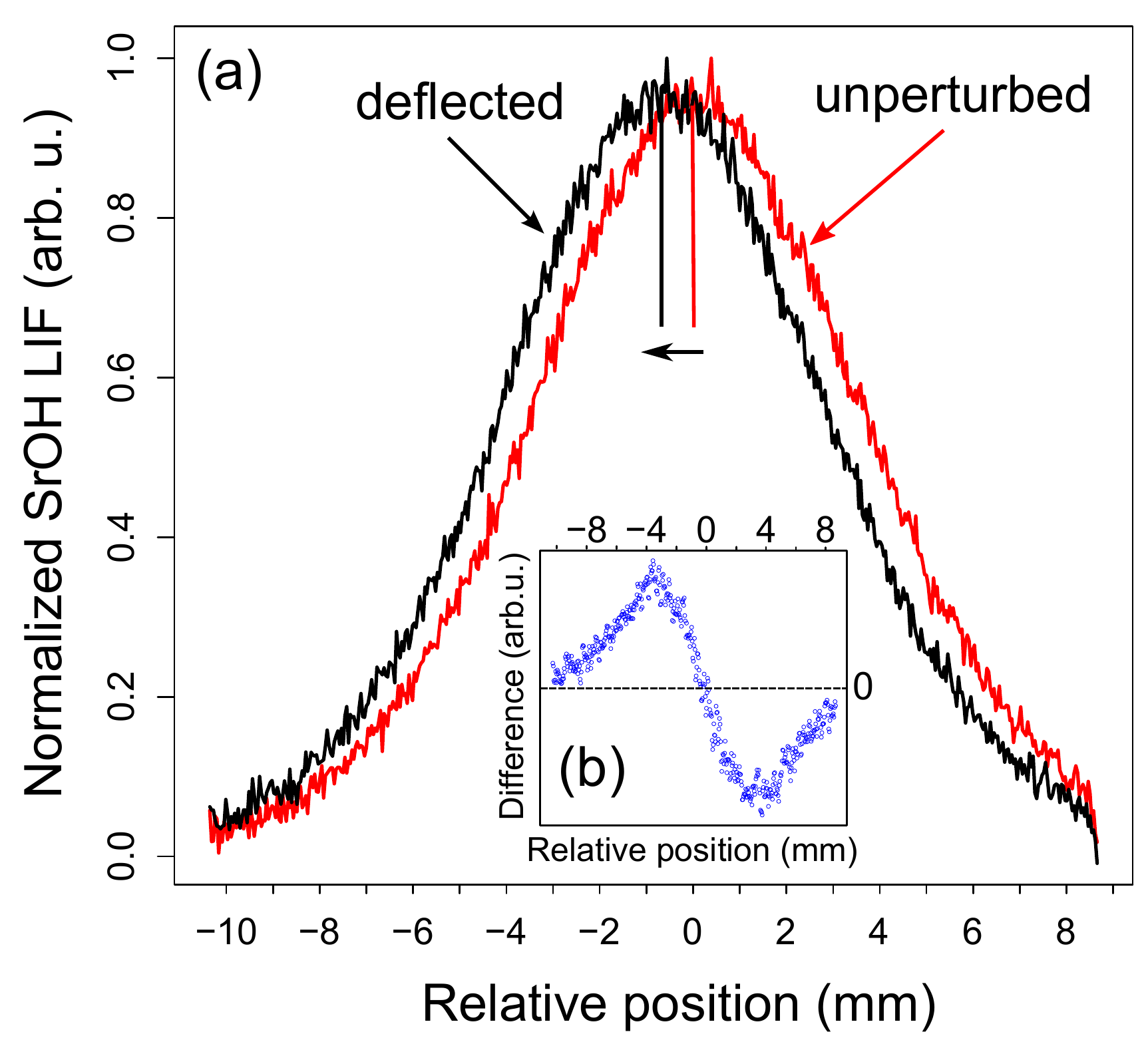} 
\par\end{centering}

\protect\protect\caption{\label{fig:Deflection-of-the-SrOH}Deflection of the SrOH beam due
to optical cycling. (a) Unperturbed spatial profile of the SrOH beam
is shown in red. Upon the application of the transverse main and repump
lasers in the interaction region the center of the beam distribution
is shifted in the negative $y$ direction (black curve). The shift
of 0.65 mm between the centers of the peaks corresponds to the scattering
of $\sim90$ photons per molecule. (b) The measured difference between
the normalized beam profiles vs position is shown. The deviation from
zero (dashed black line) indicates the deflection of the molecular
beam. }
\end{figure}

The effect of the radiation pressure force on SrOH due to photon cycling
is shown in the deflection of the molecular beam (see figure \ref{fig:Deflection-of-the-SrOH}).
In order to extract the shift per photon in our setup, we perform
cycling between (000) and (100) vibrational levels of the electronic
ground state as described in figure \ref{fig:Cycling-between-vibrational-levels}.
We measure the deflection with only the 688 nm laser applied in the
interaction region and the 631 nm clean-up beam. We observe a shift
of 0.007 mm/photon, which is consistent with the value estimated from
the travel distance and the forward velocity of the molecular beam.
The measured deflection shown in figure \ref{fig:Deflection-of-the-SrOH}(a)
corresponds to about 90 scattered photons in the interaction region
or 110 total scattered photons per molecule. From the comparison of
the unnormalized signals scaled by the in-cell absorption (which indicates
the total number of molecules produced), we determine that $\sim20\%$
of the molecules are lost during the deflection process. Using a Bernoulli
sequence to model the absorption-emission cycles \cite{di2004laser},
we estimate that the data indicate that the combined FC factor for
loss to all dark vibrational levels is $\left(3\pm1\right)\times10^{-3}$.
Employing the Sharp-Rosenstock method \cite{sharp1964franck}, we
calculate FC factors for decay from $\tilde{A}\left(000\right)$ to
$\tilde{X}\left(200\right)$ and $\tilde{X}\left(02^{0}0\right)$
to be $<1\times10^{-3}$, but vibronic coupling and anharmonic terms
in the potential could increase these decay rates \cite{fischer1984vibronic,brazier1985laser}.
In particular, the large anharmonic contribution to the bending mode
potential in SrOH \cite{presunka1995laser} could lead to enhanced
decay to the $\tilde{X}\left(02^{0}0\right)$ state. The experimental
efforts to precisely measure the FC factor for $\tilde{A}\left(000\right)\rightarrow\tilde{X}\left(02^{0}0\right)$
are currently underway in our lab.

\section{Conclusions}

By repumping out of only a single excited vibrational state, we demonstrated
cycling of $\sim110$ photons in SrOH. The $\sim90$ photons absorbed
in the interaction region lead to a $0.65\ {\rm mm}$ deflection of
a cryogenic beam of SrOH. Our estimations indicate that the molecules
lost from the photon cycle end up in the $\tilde{X}\left(02^{0}0\right)$
excited bending mode. An additional repumping laser to pump the molecules
out of the bending mode and back into the photon cycle would lead
to scattering of $\sim1,000$ photons per molecule, potentially allowing
for laser cooling of SrOH to millikelvin temperatures \cite{Shuman2010}.
Furthermore, the demonstrated optical cycling scheme opens a path
towards the use of optical bichromatic forces \cite{chieda2011prospects,aldridge2016simulations,galica2013four}
for rapid deceleration of SrOH originating from a CBGB \cite{Lu2011}
to near the capture velocity of a molecular MOT \cite{hemmerling2016CaF}.

While SrOH has a linear geometry in the vibronic ground state, it
also serves as a useful test candidate for the feasibility of laser
cooling of more complex, nonlinear molecules. Other strontium monoalkoxide
free radicals \cite{bernath1997spectroscopy} look particularly promising
for laser cooling applications. Our results corroborate previous observations
that the $\tilde{X}\rightarrow\tilde{A}$ electronic transition in
SrOH promotes a strontium-centered, nonbonding electron, leading to
highly diagonal FC factors and the dominant vibrational activity associated
with the Sr-O stretching mode \cite{brazier1985laser,brazier1986laser}.
Thus, replacing the hydrogen atom with a more complex group R (e.g.
CH\textsubscript{3} and CH\textsubscript{2}CH\textsubscript{3})
should not perturb the valence electron significantly \cite{brazier1986laser}.
SrO-R molecules share a number of properties with SrOH that are advantageous
for laser cooling, including the previously mentioned very ionic Sr-O
bond, linear local symmetry near the metal, diagonal Franck-Condon
factors, and technically accessible laser transitions \cite{brazier1986laser}.
Further work in this vein should include the evaluation of FC factors
for other vibrational modes and effects of Jahn-Teller coupling \cite{herzberg1966molecular}.

\section*{Acknowledgments}

We would like to thank D. DeMille, J. Barry, T. Steimle, and P. Bernath
for insightful discussions, as well as A. Sedlack for experimental
contributions. This work was supported by the AFOSR.

\bibliographystyle{apsrev}
\bibliography{SrOH_deflection_v2}

\end{document}